\documentclass{iau} 
\usepackage{graphicx}
\usepackage{natbib}
\usepackage{hyperref}

\title[Unseen companions in massive SB1 binaries] 
{The Nature of Unseen Companions in Massive Single-Line Spectroscopic Binaries}

\author[Hugues Sana \textit{et al.}]   
{Hugues Sana$^{1}$,
Michael Abdul-Masih$^{2}$,
Gareth Banyard$^{1}$,
Julia Bodensteiner$^{3}$,
Dominic M.\ Bowman$^{1}$,
Karan Dsilva$^{1}$,
C. Eldridge$^{4}$,
Matthias Fabry$^{1}$, 
Abigail J.\ Frost$^{1}$,
Calum Hawcroft$^{1}$,
Soetkin Janssens$^{1}$,
Laurent Mahy$^{1,5}$,
Pablo Marchant$^{1}$,
Norbert Langer$^{6,7}$,
Timothy Van Reeth$^{1}$,
Koushik Sen$^{6,7}$,
Tomer Shenar$^{1,8}$
}

\affiliation{$^1$Institute of Astronomy, KU Leuven, Celestijnenlaan 200D, 3001 Leuven, Belgium \\ $^2$email: {\tt hugues.sana@kuleuven.be}
\\[\affilskip]
$^{2}$ESO, Alonso de Cordova 3107, Vitacura,
Santiago de Chile, Chile
\\[\affilskip]
$^{3}$ESO, Karl-Schwarzschild-Strasse 2,
85748 Garching, Germany
\\[\affilskip]
$^{4}$Amateur Astronomer, 242 Randell Rd, 6125 Mardella, Western Australia
\\[\affilskip]
$^5$Royal Observatory of Belgium, Avenue Circulaire/Ringlaan 3, B-1180 Brussels, Belgium
\\[\affilskip]
$^{6}$Argelander-Institut f\"ur Astronomie, Universit\"at Bonn, Auf dem H\"ugel 71, 53121 Bonn (D)
\\[\affilskip]
$^{7}$Max-Planck-Institut f\"ur Radioastronomie, Auf dem H\"ugel 69, 53121 Bonn, Germany
\\[\affilskip]
$^8$Anton Pannekoek Institute for Astronomy, Science Park 904, 1098 XH, Amsterdam (NL)
}

\pubyear{2022}
\volume{361}  
\jname{Massive Stars Near and Far}
\editors{N.~St-Louis, J.~S.~Vink \& J.~Mackey, eds.}
\begin{document}

\maketitle

\begin{abstract}
Massive stars are predominantly found in binaries and higher order multiples. While the period and eccentricity distributions of OB stars are now well established across different metallicity regimes, the determination of mass-ratios has been mostly limited to double-lined spectroscopic binaries. As a consequence, the mass-ratio distribution remains subject to significant uncertainties. Open questions include the shape and extent of the companion mass-function towards its low-mass end and the nature of undetected companions in single-lined spectroscopic binaries.
In this contribution, we present the results of a large and systematic analysis of a sample of over 80 single-lined O-type spectroscopic binaries (SB1s) in the Milky Way and in the Large Magellanic Cloud (LMC). We report on the developed methodology, the constraints obtained on the nature of SB1 companions, the distribution of O star mass-ratios at LMC metallicity and the occurrence of quiescent OB+black hole binaries.

\end{abstract}

\firstsection 
      
\section{Introduction}

The majority of massive main-sequence OB stars are found in binary or higher order multiples \citep{KF07, SLL14, MD17}. This fact has been established beyond reasonable doubt from observational studies of various clusters and OB associations in our Galaxy \citep{Sdd12, KKL14, BSM22, BFS22} and in the Tarantula region of the Large Magellanic Cloud \citep[LMC;][]{Sdd13b,DDS15a}. Similarly, concrete observational constraints have been obtained on the distributions of orbital periods and eccentricities, again both in the Milky Way \citep{Sdd12, BGA17, BSM22}, and in the LMC \citep{AST17a, VTE21}. These studies reveal no strong differences despite the different metallicity environments. The distribution of mass-ratios, however, has remained more difficult to quantify because direct measurements of mass-ratios are mostly obtained from double-lined spectroscopic binaries (SB2s). This requires that the secondary companion is bright enough for its spectral signature to be visible in the spectrum, and that the Doppler separation between the components is large enough for the spectral lines to separate enough. In some cases, e.g., when temperature differences are large enough to result in different ionisation balances, radial velocities (RVs) of the components can be measured from lines unaffected by blending, but this is not the case in the majority of the stars, and even a small degree of contamination can impact the RV measurements \citep{FHF21, BSW21, BSM22}.

Focusing on the Galactic population within a few kiloparsecs, about one third of the known O-type spectroscopic binaries  show the spectroscopic signature of only one of the binary components \citep{TBN21, MSS22}, i.e. they are single-lined spectroscopic binaries (SB1s). This SB1 fraction increases to over 50\% in LMC studies \citep{AST17a}, most likely owing to the lower data quality given stars are about 10 times further away. Pending further information, the precise nature of the unseen companions in SB1 systems remains uncertain. One may expect that the majority are `normal' lower-mass stars whose luminosities are too low for their spectral lines to peak out of the noise in individual-epoch spectroscopy, or whose orbital periods are too large for their spectral lines to sufficiently separate from those of the brighter component. Alternatively, such systems might also harbour binary interaction products where the initially more massive star (mass-donor) has been stripped of (most of) its hydrogen envelope through mass transfer \citep{Pac67, Pols91, ABB20, BSM20a, SBA20a, EBQ21}. SB1s may also be the hiding place of compact objects where the unseen companions is either a black hole (BH) or a neutron star \citep{LSS20}. 

Identifying the nature of unseen companions is important for a number of science cases:
\begin{enumerate}
\item[{\bf Multiplicity properties:}] present observational constraints of the mass-ratio distributions of unevolved binaries rely on extrapolation towards the lower-mass-end as low mass-ratios systems ($M_2/M_1 \lesssim 0.4 $) are rarely detected as SB2. The initial mass-ratio distribution is, however, critical for binary evolution computations as the mass ratio is a fundamental parameter in defining the outcome of the interaction;\\
\item[{\bf Binary evolution:}] a frequent outcome for systems that survive a phase of mass-transfer is that the mass-donor is stripped of (most of) its hydrogen envelope while the mass accretor is (probably) a rapidly rotating OB star. Stripped helium stars are candidate hydrogen-poor supernova progenitors \citep{L12}. Furthermore, the distribution of orbital properties and the chemical composition of stars in these systems are fundamental benchmarks to confront the outcome of mass-transfer models, yet these OB+He systems likely appear as SB1 in the optical;\\
\item[{\bf Compact objects:}] recent binary population synthesis simulations predict that about 2\% of O-type stars should be in a binary with a BH companion \citep{LSS20}. Focusing on SB1 systems, this fraction might rise to 8\%. Predicted orbital periods for these OB+BH systems range from about a week to a couple of years. This makes spectroscopy an ideal tool to search for these (most likely) X-ray quiet BH binaries. Alternatively, neutron stars can also be present around lower-mass O stars or around B-type stars. The orbital properties of these systems retain pristine information about the collapse of the BH/NS progenitors (Marchant et al., this volume), information that might be distorted or  lost once these systems evolve into X-ray binaries.

\end{enumerate}
 
\section{Method}
\begin{figure}[ht!]
\begin{center}
\includegraphics[width=11cm]{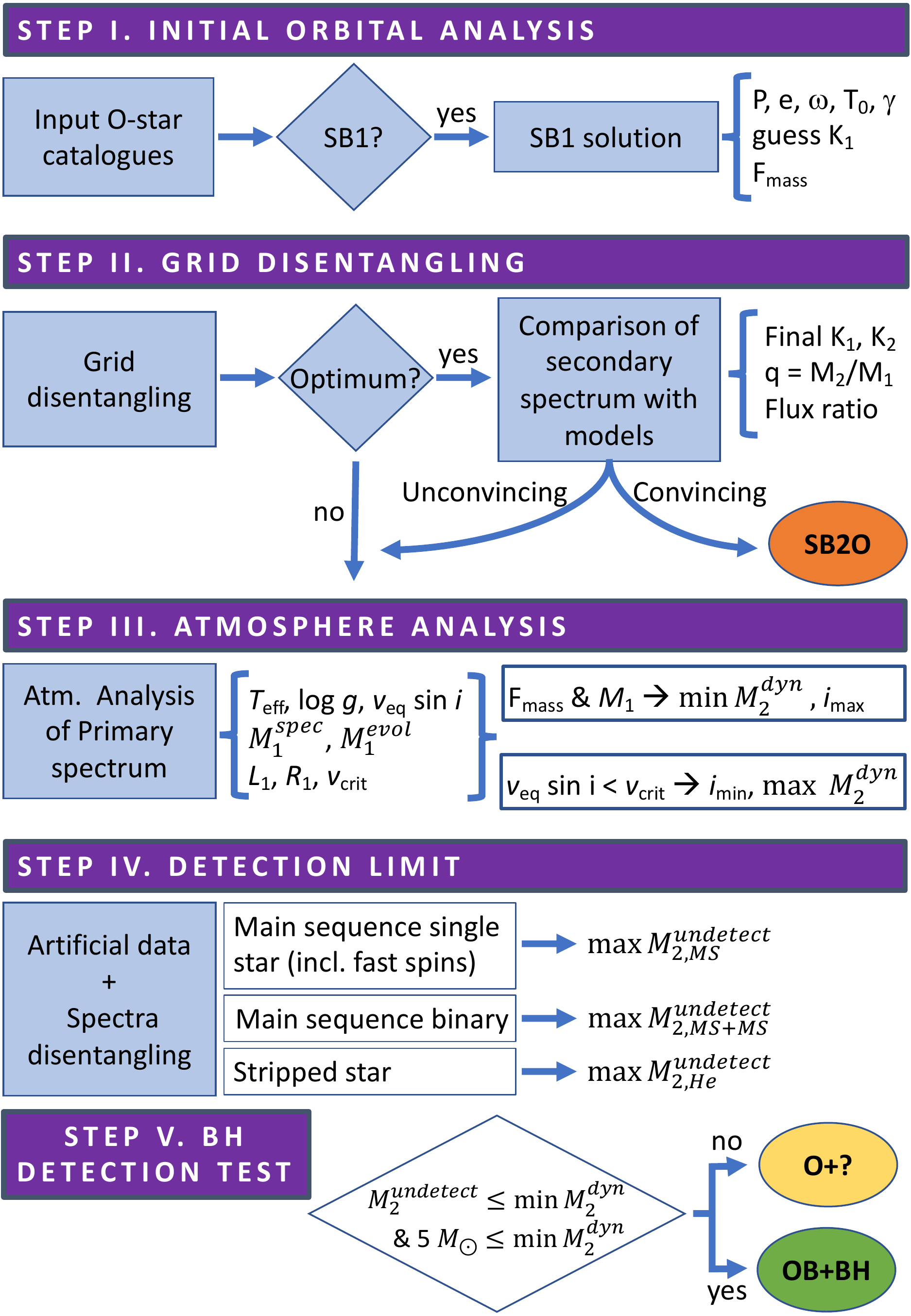}

 \caption{Workflow for identifying the nature of unseen companions in SB1 OB binaries.}
   \label{f:BHsearch}
\end{center}
\end{figure}

We worked under the assumption that constraints on the mass, luminosity and spectral properties of the unseen object help us to better constrain its nature and physical properties. We relied on four main steps, described below and summarised in Fig.~\ref{f:BHsearch}: 
\begin{enumerate}
\item[{\bf $\bullet$ SB1 orbital analysis:}] the SB1 orbital solution of the visible component provides reliable constraints on the orbital periods ($P$), ephemeris and geometry of the orbit. A first estimate of the semi-amplitude of the RV curve of the visible component ($K_1$) is obtained but does not need to be trusted, as even a small contamination of its spectral lines with that of a faint unidentified companion may bias the measurements, leading to an underestimation of $K_1$;

\item[{\bf $\bullet$ Grid-based spectral disentangling:}] the  well-constrained orbital properties are kept fixed and the disentangling is performed on a 2D grid along $K_1$ and $K_2$. We tested both Fourier-based and shift-and-add  disentangling methods \citep{H95, MME98, GL06}, with mostly similar results. In some cases where the semi-amplitude of the RV curve of the unseen companion ($K_2$) is close to null \citep[e.g.,][]{BSM20a}, the shift-and-add method has been yielding slightly more consistent results but the fundamental numerical reasons have not yet been identified; 

\item[{\bf $\bullet$ Atmosphere analysis:}] a comparison of the disentangled spectra with atmosphere models allows us (i) to decide whether the stellar spectral signature of the companion can be identified, (ii) to estimate the optical light ratio, and (iii) to constrain the stellar properties of the optically bright component, including its spectroscopic and evolutionary masses, projected rotation rate and constraints on the orbital inclination. This provides us with a possible dynamical mass range for the unseen companion;

\item[{\bf $\bullet$ Detection limits:}] if no stellar companion is identified by spectral disentangling, we  perform simulations to estimate the maximum mass at which the spectral signature of a non-degenerate companion would remain undetected through our disentangling analysis. The comparison between the outcome of the simulations and the dynamical mass range of the unseen companion allows us to put additional constraints on the companion mass. Specifically, if the minimum mass of the unseen companion is larger than 5Msun and non-degenerate companions within the acceptable mass range should have led to a detection with spectral disentangling, then one is left to accept that the system is likely an OB+BH binary. In establishing the detection limits, it is important to consider companions of different natures. Stripped He star companions have indeed led to a number of false detections \citep[e.g.,][]{BSM20a, EBQ21, FBR22} because of their significantly different mass-optical brightness relation compared to single main-sequence companions \citep{GdG18}. An unseen component which is in fact a very short period near-equal mass binary more easily eludes detection compared to a main sequence star of the same total mass. Similarly, large projected rotation rates decrease the detectability of the unseen companion as its spectral lines are smeared out.
\end{enumerate}

\section{Sample}
Our project focused on two samples of O-type SB1 systems:
\begin{enumerate}
\item[{\bf $\bullet$ 32 Galactic systems}] identified through literature search and the Galactic O star Spectroscopic Catalogue \citep{MSA16}: we used high spectral resolving power ($R$), high-$S/N$ spectra obtained from archival and new observations from ESO, Mercator/HERMES \citep{RvH11} and SALT facilities. Of interest, Cyg X-1, known to host a stellar-mass BH, was included in the sample as it met our selection criteria (O-type SB1 + availability of sufficient spectroscopic data); see \citet{MSS22}.

\item[{\bf $\bullet$ 51 systems in the Tarantula region of the LMC}] identified by the VLT-FLAMES Tarantula Survey \citep[VFTS, ][]{ETH11a}. These were monitored for 32 spectroscopic epochs by the Tarantula Massive Binary Monitoring \citep[TMBM, ][]{AST17a} using the LR02 setup of the VLT/FLAMES multi-object spectrograph and yielding a $R\sim7000$ from about 3950 to 4550~\AA; see \citet{SSM22a,SSM22}.
\end{enumerate}
 Our search for compact objects has also been extended to Galactic and LMC B-type SB1 systems. Preliminary results are presented in Banyard et al.\ and Villase\~nor et al.\ in this volume; see also \citet{BSM22, VTE21}.

\section{Results}
Spectral disentangling revealed the signature of non-degenerate stellar companions in over 50\% of the systems considered.  Future work, including better data, is required to confirm the evolutionary status of these systems, albeit most are likely low mass-ratio binaries but some might be binary interaction products; see \citet{MSS22, SSM22a, SSM22} for further details. Our results double the number of TMBM O-type binaries with measured mass-ratios. Combined with the analysis of previously known SB2 systems \citep{MSA20, MAS20}, this gives us direct access to the largest sets of mass-ratio measurements available so far for an O+OB binary population at sub-solar metallicity. Bayesian modelling of the measurements, including correction for observational biases, yields a flat mass-ratio distribution \citep[$f_q \propto q^k$, with $k = 0.2 \pm 0.2$;][]{SSM22a}. This value is significantly different than obtained by a previous attempt based on the VFTS data ($k = -1.0 \pm 0.4$), confirming that the modelling of the amplitude of the RV variations offers low quality diagnostics on the mass-ratio distribution \citep{Sdd13b}. Of interest, the mass-ratio distribution of O stars in the Tarantula region seem compatible with that derived for O stars in  Milky Way clusters and OB associations ($k = -0.1 \pm 0.6$) suggesting no significant impact of the metallicity on the (still uncertain) pairing mechanism of massive binaries.
 
Among systems in which the signature of the companion cannot be identified, three systems in the Milky Way and three in the TMBM samples meet our requirements for OB+BH systems: (i) the companion mass is larger than $5~M_{\odot}$ and, (ii) according to our simulations, a non-degenerate component with such high mass should have been detected given our data quality and orbital sampling, and yet it is not. In doing so, we rediscovered the OB+BH nature of Cyg X-1, which builds some confidence in the ability of our method to isolate good OB+BH candidates. Among the identified systems, the two best cases are HD~130298 in the Milky Way ($P=14.6$~d, $e=0.47$)  and VFTS~243 in the LMC ($P=10.4$~d, $e<0.03$). Of interest, both have very different eccentricity properties. The negligible eccentricity in VFTS~243 suggests that the dying primary collapsed into a BH without losing much mass ($\Delta M< 1.1~M_\odot$), thus likely without an associated supernova (SN) explosion. The formation scenario in HD~130298 is different and the system might have lost about 25\% of the system total mass prior to SN explosion, i.e. an estimated $8~M_\odot$. This  suggests a significant mass loss and a possible BH formation through fallback.

\section{Conclusions}
We have presented a systematic analysis of a large sample of SB1 systems to identify the nature of their unseen companions. We have shown that grid disentangling along the $K_1$ and $K_2$ axes offers significant diagnostic power to extract the optical spectral signature of a faint stellar companion, which is otherwise lost in the noise of individual observational epochs. Simulations of artificial data sets show that signatures of companions representing only 0.3\% of the total flux can be retrieved pending sufficiently high-quality time series and that hydrogen lines are better suited for faint objects than metallic lines. We also showed that it is easier to detect the presence a faint stellar companion than to precisely measure its RV-curve semi-amplitude $K_2$. Simulations of non-idealised data and realistic systems have further shown that companions contributing just 1\% are easily retrieved. The lowest mass-ratio new-SB2 system revealed by our analysis has $q\sim0.15$. 
The results from the TMBM sample in the LMC point toward a metallicity independent, (nearly) flat mass-ratio distribution for O-type binaries. Further extension towards lower metallicity is desirable, unfortunately current and future surveys do not plan enough multi-epoch coverage to reach the needed sampling and combined $S/N$. The presented method properly identified Cyg X-1 as a OB+BH system and revealed at least five new OB+BH (strong candidate) systems. We briefly discussed how the orbital properties help constraining the collapse physics.  
 
\section*{Acknowledgements}
The research leading to these results has received funding from the European Research Council (ERC) under the European Union's Horizon 2020 research and innovation programme (grant agreement number 772225: MULTIPLES), as well as under the Marie Skłodowska-Curie grant agreement No 101024605. The KU Leuven team is grateful to the FWO for multiple support through grants G0F8H6N, 11E1721N, 1286521N, 12ZB620N, 12ZY520N and the Belgian Federal Science Policy Office (BELSPO) in the framework of the PRODEX Programme.





\begin{thebibliography}{35}

\bibitem[{{Abdul-Masih} {et~al.}(2020){Abdul-Masih}, Banyard, Bodensteiner,
  Bordier, Bowman, Dsilva, Fabry, Hawcroft, Mahy, Marchant, Raskin, Reggiani,
  Shenar, Tkachenko, Van~Winckel, Vermeylen, \& Sana}]{ABB20}
{Abdul-Masih}, M., {et~al.} 2020, Nature, 580, E11

\bibitem[{Almeida {et~al.}(2017)Almeida, Sana, Taylor, Barb{\'a}, Bonanos,
  Crowther, Damineli, {de Koter}, {de Mink}, Evans, Gieles, Grin,
  {H{\'e}nault-Brunet}, Langer, Lennon, Lockwood, Ma{\'i}z~Apell{\'a}niz,
  Moffat, Neijssel, Norman, {Ram{\'i}rez-Agudelo}, Richardson, Schootemeijer,
  Shenar, Soszy{\'n}ski, Tramper, \& Vink}]{AST17a}
Almeida, L.~A., {et~al.} 2017, A\&A, 598, A84

\bibitem[{Banyard {et~al.}(2022)Banyard, Sana, Mahy, Bodensteiner,
  Villase{\~n}or, \& Evans}]{BSM22}
Banyard, G., {et~al.}  2022, A\&A, 658, A69

\bibitem[{Barb{\'a} {et~al.}(2017)Barb{\'a}, Gamen, Arias, \& Morrell}]{BGA17}
Barb{\'a}, R.~H., Gamen, R., Arias, J.~I., \& Morrell, N.~I. 2017, in {{IAU Symposium}} Vol. 329, 
  89

\bibitem[{Bodensteiner {et~al.}(2020)Bodensteiner, Shenar, Mahy, Fabry,
  Marchant, {Abdul-Masih}, Banyard, Bowman, Dsilva, Frost, Hawcroft, Reggiani,
  \& Sana}]{BSM20a}
Bodensteiner, J., {et~al.} 2020, A\&A, 641, A43

\bibitem[{Bodensteiner {et~al.}(2021)Bodensteiner, Sana, Wang, Langer, Mahy,
  Banyard, {de Koter}, {de Mink}, Evans, G{\"o}tberg, Patrick, Schneider, \&
  Tramper}]{BSW21}
---. 2021, A\&A, 652, A70

\bibitem[{Bordier {et~al.}(2022)Bordier, Frost, Sana, Reggiani, M{\'e}rand,
  Rainot, {Ram{\'i}rez-Tannus}, \& {de Wit}}]{BFS22}
Bordier, E., {et al.} 2022, A\&A, 663, A26

\bibitem[{Dunstall {et~al.}(2015)Dunstall, Dufton, Sana, Evans, Howarth,
  {Sim{\'o}n-D{\'i}az}, {de Mink}, Langer, Ma{\'i}z~Apell{\'a}niz, \&
  Taylor}]{DDS15a}
Dunstall, P.~R., {et~al.} 2015, A\&A, 580, A93

\bibitem[{{El-Badry} \& {Quataert}(2021)}]{EBQ21}
{El-Badry}, K., \& {Quataert}, E. 2021, MNRAS, 502, 3436

\bibitem[{Evans {et~al.}(2011)Evans, Taylor, {H{\'e}nault-Brunet}, Sana, {de
  Koter}, {Sim{\'o}n-D{\'i}az}, Carraro, Bagnoli, Bastian, Bestenlehner,
  Bonanos, Bressert, Brott, Campbell, Cantiello, Clark, Costa, Crowther, {de
  Mink}, Doran, Dufton, Dunstall, Friedrich, Garcia, Gieles, Gr{\"a}fener,
  Herrero, Howarth, Izzard, Langer, Lennon, Ma{\'i}z~Apell{\'a}niz, Markova,
  Najarro, Puls, Ramirez, {Sab{\'i}n-Sanjuli{\'a}n}, Smartt, Stroud, {van
  Loon}, Vink, \& Walborn}]{ETH11a}
Evans, C.~J., {et~al.} 2011, A\&A, 530, A108

\bibitem[{Fabry {et~al.}(2021)Fabry, Hawcroft, Frost, Mahy, Marchant,
  Le~Bouquin, \& Sana}]{FHF21}
Fabry, M., {et~al.} 2021, A\&A, 651, A119

\bibitem[{Frost {et~al.}(2022)Frost, et al. }]{FBR22}
Frost, A.~J., {et~al.} 2022, A\&A, 659, A3

\bibitem[{Gonz{\'a}lez \& Levato(2006)}]{GL06}
Gonz{\'a}lez, J.~F., \& Levato, H. 2006, A\&A, 448, 283

\bibitem[{G{\"o}tberg {et~al.}(2018)G{\"o}tberg, {de Mink}, Groh, Kupfer,
  Crowther, Zapartas, \& Renzo}]{GdG18}
G{\"o}tberg, Y., {et~al.} 2018, A\&A 615, A78

\bibitem[{Hadrava(1995)}]{H95}
Hadrava, P. 1995, A\&A Supplement, v.114, p.393, 114, 393

\bibitem[{Kobulnicky \& Fryer(2007)}]{KF07}
Kobulnicky, H.~A., \& Fryer, C.~L. 2007, ApJ, 670, 747

\bibitem[{Kobulnicky {et~al.}(2014)Kobulnicky, Kiminki, Lundquist, Burke,
  Chapman, Keller, Lester, Rolen, Topel, Bhattacharjee, Smullen,
  Vargas~{\'A}lvarez, Runnoe, Dale, \& Brotherton}]{KKL14}
Kobulnicky, H.~A., {et~al.} 2014, APJSS, 213, 34

\bibitem[{Langer(2012)}]{L12}
Langer, N. 2012, ARA\&A, 50, 107

\bibitem[{Langer {et~al.}(2020)Langer, Sch{\"u}rmann, Stoll, Marchant, Lennon,
  Mahy, {de Mink}, Quast, Riedel, Sana, Schneider, Schootemeijer, Wang,
  Almeida, Bestenlehner, Bodensteiner, Castro, Clark, Crowther, Dufton, Evans,
  Fossati, Gr{\"a}fener, Grassitelli, Grin, Hastings, Herrero, {de Koter},
  Menon, Patrick, Puls, Renzo, Sander, Schneider, Sen, Shenar,
  {Sim{\'o}n-D{\'i}as}, Tauris, Tramper, Vink, \& Xu}]{LSS20}
Langer, N., {et~al.} 2020, A\&A, 638, A39

\bibitem[{Mahy {et~al.}(2022)Mahy, Sana, \& Shenar}]{MSS22}
Mahy, L., Sana, H., \& Shenar, T. 2022, A\&A, 664, 159

\bibitem[{Mahy {et~al.}(2020{\natexlab{a}})Mahy, Sana, {Abdul-Masih}, Almeida,
  Langer, Shenar, {de Koter}, {de Mink}, {de Wit}, Grin, Evans, Moffat,
  Schneider, Barb{\'a}, Clark, Crowther, Gr{\"a}fener, Lennon, Tramper, \&
  Vink}]{MSA20}
Mahy, L., {et~al.} 2020{\natexlab{a}}, A\&A, 634, A118

\bibitem[{Mahy {et~al.}(2020{\natexlab{b}})Mahy, Almeida, Sana, Clark, {de
  Koter}, {de Mink}, Evans, Grin, Langer, Moffat, Schneider, Shenar, \&
  Tramper}]{MAS20}
---. 2020{\natexlab{b}}, A\&A, 634, A119

\bibitem[{Ma{\'i}z~Apell{\'a}niz {et~al.}(2016)Ma{\'i}z~Apell{\'a}niz, Sota,
  Arias, Barb{\'a}, Walborn, {Sim{\'o}n-D{\'i}az}, Negueruela, Marco, Le{\~a}o,
  Herrero, Gamen, \& Alfaro}]{MSA16}
Ma{\'i}z~Apell{\'a}niz, J., {et~al.} 2016, APJSS, 224, 4

\bibitem[{Marchenko {et~al.}(1998)Marchenko, Moffat, \& Eenens}]{MME98}
Marchenko, S.~V., Moffat, A. F.~J., \& Eenens, P. R.~J. 1998, PASP, 110, 1416

\bibitem[{Moe \& Di~Stefano(2017)}]{MD17}
Moe, M., \& Di~Stefano, R. 2017, APJS,
  230, 15

\bibitem[{{Paczy{\'n}ski}(1967)}]{Pac67}
{Paczy{\'n}ski}, B. 1967, Acta Astron., 17, 193

\bibitem[{{Pols} {et~al.}(1991){Pols}, {Cote}, {Waters}, \& {Heise}}]{Pols91}
{Pols}, O.~R., {Cote}, J., {Waters}, L.~B.~F.~M., \& {Heise}, J. 1991, A\&A,
  241, 419

\bibitem[{Raskin {et~al.}(2011)Raskin, {van Winckel}, Hensberge, Jorissen,
  Lehmann, Waelkens, Avila, {de Cuyper}, Degroote, Dubosson, Dumortier,
  Fr{\'e}mat, Laux, Michaud, Morren, Perez~Padilla, Pessemier, Prins, Smolders,
  {van Eck}, \& Winkler}]{RvH11}
Raskin, G., {et~al.} 2011, A\&A, 526, A69

\bibitem[{Sana {et~al.}(2012)Sana, {de Mink}, {de Koter}, Langer, Evans,
  Gieles, Gosset, Izzard, Le~Bouquin, \& Schneider}]{Sdd12}
Sana, H., {et~al.} 2012, Science, 337, 444

\bibitem[{Sana {et~al.}(2013)Sana, {de Koter}, {de Mink}, Dunstall, Evans,
  {H{\'e}nault-Brunet}, Ma{\'i}z~Apell{\'a}niz, {Ram{\'i}rez-Agudelo}, Taylor,
  Walborn, Clark, Crowther, Herrero, Gieles, Langer, Lennon, \& Vink}]{Sdd13b}
---. 2013, A\&A, 550, A107

\bibitem[{Sana {et~al.}(2014)Sana, Le~Bouquin, Lacour, Berger, Duvert, Gauchet,
  Norris, Olofsson, Pickel, Zins, Absil, {de Koter}, Kratter, Schnurr, \&
  Zinnecker}]{SLL14}
---. 2014, APJS, 215, 15

\bibitem[{Shenar {et~al.}(2020)Shenar, Bodensteiner, {Abdul-Masih}, Fabry,
  Mahy, Marchant, Banyard, Bowman, Dsilva, Hawcroft, Reggiani, \&
  Sana}]{SBA20a}
Shenar, T., {et~al.} 2020, A\&A, 639, L6

\bibitem[{Shenar {et~al.}(2022{\natexlab{a}})Shenar, Sana, Mahy, \& {et
  al.}}]{SSM22a}
---.  2022{\natexlab{a}}, A\&A

\bibitem[{Shenar {et~al.}(2022{\natexlab{b}})Shenar, Sana, Mahy, \& {et
  al.}}]{SSM22}
---. 2022{\natexlab{b}}, Nature Astronomy (https://doi.org/10.1038/s41550-022-01730-y)

\bibitem[{Trigueros~P{\'a}ez {et~al.}(2021)Trigueros~P{\'a}ez, Barb{\'a},
  Negueruela, Ma{\'i}z~Apell{\'a}niz, {Sim{\'o}n-D{\'i}az}, \& Holgado}]{TBN21}
Trigueros~P{\'a}ez, E., {et~al.} 2021,
  A\&A, 655, A4

\bibitem[{Villase{\~n}or {et~al.}(2021)Villase{\~n}or, Taylor, Evans,
  {Ram{\'i}rez-Agudelo}, Sana, Almeida, {de Mink}, Dufton, \& Langer}]{VTE21}
Villase{\~n}or, J.~I., {et~al.} 2021, MNRAS, 507, 5348

\end{thebibliography}
\end{document}